%% file: ml4ps_cameraready.tex
\documentclass{article}
\PassOptionsToPackage{numbers,sort&compress}{natbib}
\usepackage[final,dblblindworkshop]{neurips_2026} 

\usepackage[utf8]{inputenc}
\usepackage[T1]{fontenc}
\usepackage{hyperref}
\usepackage{url}
\usepackage{booktabs}
\usepackage{capt-of}   
\usepackage{amsfonts}
\usepackage{amsmath}
\usepackage{nicefrac}
\usepackage{microtype}
\usepackage{xcolor}
\usepackage{graphicx}
\usepackage{titlesec}
\graphicspath{{figures/}}
\titlespacing{\section}{0pt}{0pt}{0pt}
\titlespacing{\subsection}{0pt}{0pt}{0pt}
\titlespacing{\paragraph}{0pt}{0pt}{0pt}

\title{Toward a Survey-Agnostic Representation for Strong Gravitational Lensing}

\author{%
  Sreevani Jarugula \\
  Computational Science and AI Directorate\\
  Fermilab\\
  Batavia, IL 60510 \\
  \texttt{jarugula@fnal.gov} \\
  \And
    Aleksandra \'Ciprijanovi\'c \\
  Scientific Computing Division\\ 
  Fermi National Accelerator Laboratory, Batavia, IL 60510 \\
  Department of Astronomy and Astrophysics\\
  University of Chicago, Chicago, IL 60637\\
  NSF-Simons AI Institute for the Sky (SkAI)\\
  172 E. Chestnut St., Chicago, IL 60611\\
  \texttt{aleksand@fnal.gov}
}

\begin{document}
\maketitle

\begin{abstract}
\vspace{-0.2cm}
Modern and upcoming astronomy surveys are imaging billions of galaxies, among them galaxy-galaxy strong gravitational lenses, which are powerful probes of astrophysics and cosmology. Combining lens observations across surveys improves model parameter inference, but each telescope is limited by its optics and observing conditions, which blur the image by convolving with a instrument specific point spread function (PSF), pixel sampling, and noise. Hence, a model trained on one telescope does not directly apply to another. We present a model that can ingest images across telescopes, recovering the sharp, deconvolved image and the Einstein radius at each survey's native resolution. In this transformer architecture, positions are expressed in physical units (arcseconds) rather than pixel indices, and PSF learns the deconvolution through cross-attention with the image. When applied to simulated lenses from five different surveys spanning a $5.3\times$ range of pixel scale, our model recovers the Einstein radius with RMSE of $1.3$--$3.7\%$ and reduces the high spatial frequency reconstruction error $10.9$--$41.7\times$ relative to the blurred input. The inclusion of PSF cross-attention improves high frequency recovery by $1.6$--$10\times$, highlighting the importance of our method for image deconvolution.
\end{abstract}

\section{Introduction}
\label{sec:intro}

Current and upcoming astronomical surveys are imaging the sky at the petabyte scale and will catalog billions of galaxies over the next decade. Wide-field ground-based surveys such as Rubin's Legacy Survey of Space and Time (LSST)~\cite{ivezic2019lsst} and the Dark Energy Survey (DES)~\cite{des2016overview} trade angular resolution for coverage, while space-based instruments such as the Hubble Space Telescope (HST)~\cite{hst24}, Euclid~\cite{euclid2025overview}, and
Roman~\cite{schlieder24} approach the diffraction limit over smaller
fields. A joint analysis of data from these surveys holds more information
than any one survey's image alone but, this is difficult to achieve because each telescope blurs the sky with its own point spread function (PSF) and samples on its own pixel grid. Recovering the intrinsic signal (deconvolution) is therefore crucial for measuring galaxy shapes, fluxes, and substructure without
bias, and for comparing images taken at different resolutions. However, deconvolution is an ill-posed problem and classical methods such as Wiener filtering~\citep{wiener1949extrapolation} and Richardson--Lucy iteration~\citep{richardson1972, lucy1974}, solve this analytically but require accurate knowledge of the PSF and the noise spectrum and needs regularization. Neural networks offer faster and more flexible alternatives to conventional algorithms for deconvolution of astronomical data. However, these models are mainly trained for a specific telescope and the models which learn the astrophysical signal along with telescope/survey properties such as image PSF, pixel scale, and noise, can fail in two ways: it does not transfer from one survey to another, and it does not transfer from simulation to real observations~\citep[e.g.][]{ciprijanovic23, swierc2024domain}. We develop a neural network that decouples the effect of PSF and pixel scale by leveraging their information, where the reconstruction utilizes the full PSF image through cross-attention resulting in a learned representation of the intrinsic signal that supports downstream tasks such as inference. We apply this network to strong gravitational lensing, in which a foreground mass deflects the light of a background galaxy. Lensing is a powerful probe of dark matter and dark
energy~\citep{treu2010strong}, and Rubin, Euclid, and Roman will discover thousands of new lenses over the next decade, motivating our analysis. Although we train the model on single-band data without survey noise and assuming a Gaussian PSF, separating the instrument nuisances from the signal is a step toward applying simulation-trained models to the real observations.

\paragraph{Related work:} \hspace{0.1cm} Classical deconvolution by Wiener
filtering and Richardson--Lucy iteration methods
remains the benchmark, but require a known PSF and noise spectrum with careful tuning of the regularization. Neural networks have since improved both quality and
speed: U-Net architectures have been successfully implemented for deconvolution of galaxy images along with custom loss functions to preserve galaxy shapes~\citep {sureau2020spacevariant, nammour2022shapenet}.  Learnlets model~\citep{ramzi23} combines wavelets and  U-Nets, which have been used for deconvolution of Hubble Space Telescope (HST) data~\citep{akhaury22}. Score based methods have shown promising results for deconvolution and lens reconstruction~\citep{adam25,legin25}.
However, these models are mainly developed and trained for a specific telescope. Cross-survey information has been exploited by \citet{akhaury2025multiband} to sharpen Rubin images using overlapping Euclid data, demonstrating the value of joint analysis. However, this still requires the high-resolution survey during inference. Closest to our
approach, \citet{patel25} condition neural posterior estimation on a learned PSF embedding while \citet{ni24} learns the sharp image by using the PSF for blurring in the loss function. In both cases the PSF is not directly encoded in the reconstruction and is compressed to an embedding for parameter inference, or serves only to degrade the output for training. In our model the reconstruction utilizes the full PSF image through cross-attention, and a single encoder therefore deconvolves images from any survey at their native pixel scale, given the measured PSF. 

\section{Data and Methods}
\label{sec:methods}

\begin{figure}[t]
\centering
\begin{minipage}[c]{0.6\textwidth}\centering
  \includegraphics[width=0.95\linewidth]{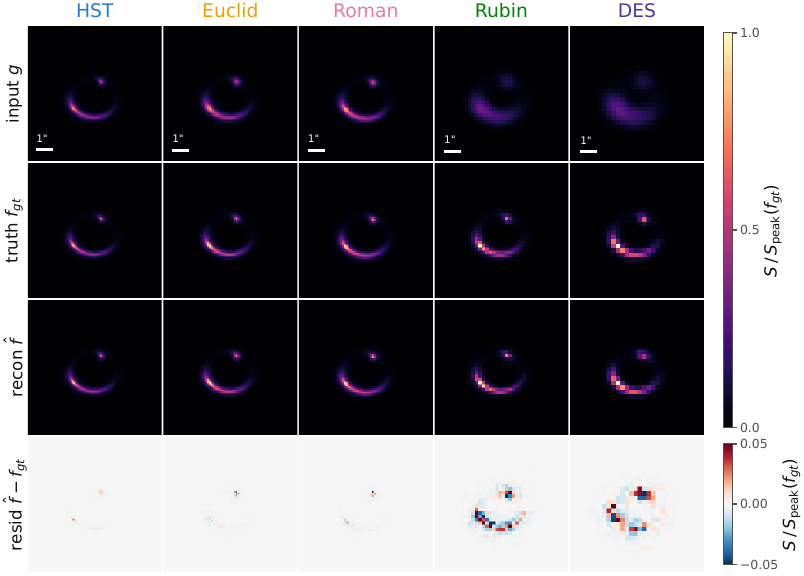}
\end{minipage}\hfill
\begin{minipage}[c]{0.4\textwidth}\centering
  \footnotesize
  \captionof{table}{Survey configurations for a $12''$ field of view.}
  \label{tab:surveys}
  \setlength{\tabcolsep}{3.5pt}
  \begin{tabular}{lccccc}
  \toprule
                    & HST  & Euclid & Roman & Rubin & DES   \\
  \midrule
  scale ($''$/px)   & 0.05 & 0.10   & 0.11  & 0.20  & 0.263 \\
  FWHM ($''$)       & 0.10 & 0.16   & 0.12  & 0.70  & 0.90  \\
  PSF $\sigma$ (px) & 0.85 & 0.68   & 0.46  & 1.49  & 1.45  \\
  grid (px)         & 240  & 120    & 109   & 60    & 46    \\
  \bottomrule
  \end{tabular}
\end{minipage}
\caption{The same lens observed by five telescopes and processed at native resolution. Columns are surveys ordered by pixel scale
(Table~\ref{tab:surveys}); rows are the input $g$, the truth $f_{gt}$, the
reconstruction $\hat f$ and the residual $\hat f - f_{gt}$, on a $\pm0.05$
scale of the truth peak. Each column is normalised by its own truth peak. The $1''$ bar is identical sky in
every column.}
\label{fig:hero}
\end{figure}

\paragraph{Data.}~
Lens images are generated with \texttt{lenstronomy}
\citep{birrer2018lenstronomy} in a fixed $12''$ field of view with arbitrary flux units: an elliptical power-law mass with external shear lensing a Sérsic source. The observational parameters are randomly drawn from
continuous priors that include all five surveys we want to evaluate on. 
To generate an image, the forward model renders the sharp signal on a fine grid, convolves it with the randomly chosen PSF, and area-averages to the randomly chosen target pixel scale. We train on $10{,}000$ images 
and for testing, we draw $200$ lens images once and render each through five surveys: HST, Euclid, Roman, Rubin and DES (Table~\ref{tab:surveys}). Hence, the differences between the five test sets are only due to telescope properties, not the lens sample.
Train and test images are noise-free, so a cross-survey difference measures the effect of the PSF
and pixel scale alone.

\paragraph{Architecture.}~
Every telescope blurs an observed sharp signal $f$ by convolving with a known PSF $h$, to create a blurred image $g$ as: $g = f * h$. The sharp image $f$ is extracted through deconvolution. Our network takes three inputs: The native pixel scale of the survey $s$ (Table~\ref{tab:surveys}), the image $g$, and PSF $h$, both at the same native pixel grid. The image and PSF are divided
into fixed $4\times4$ pixel patches and projected to $d=128$. 
Invariance to pixel size is built with positional embedding. Each patch center becomes a physical offset from the lens center,
$(x, y) = (i\,s/\theta_{\mathrm{ref}},\; j\,s/\theta_{\mathrm{ref}})$ with
$\theta_{\mathrm{ref}} = 0.45''$ where $\{i,j\}$ are the patch coordinates. 2D rotary position embeddings (RoPE) 
\citep{su2021roformer,heo2024rope} are applied in this frame. RoPE makes the attention logit depend only on the separation: the same angular offset produces the same rotation for every instrument, and, since convolution
couples $f$ and $h$ only through coordinate differences, image-PSF cross-attention 
modulated by $x_i - x_p$ carries exactly the displacement structure
deconvolution requires. The encoder stacks four identical transformer blocks, each applying three
attention operations: self-attention over image tokens, self-attention over
PSF tokens, and cross-attention in which image tokens query the PSF tokens. Two heads read the embedding to perform: 1) deconvolution and 2) regression of Einstein Radius ($\theta_E$). 
The model has $1.91$M parameters and is trained for $80$ epochs. We use the AdamW~\citep{loshchilov2019adamw} optimizer at a learning rate of $3\times10^{-4}$ with cosine decay and batch size $8$. Training takes ${\sim}2.25$ hours on a A100 GPU. The code used in this analysis is available in the GitHub repository\footnote{\url{https://github.com/jsv1206/survey-invariant-lensnn}}.

\paragraph{Objective:}~
Training is supervised, guided by three main objectives: reconstruction, reblurring, and regression. With $f_{gt}$ being
the sharp image truth, and $\hat f$ its reconstruction, the training objective is:
{\footnotesize
\begin{equation}
\mathcal{L}
 = \underbrace{\|\hat f - f_{gt}\|_1
   + \lambda_a\big(1 - \mathrm{SSIM}(\hat f, f_{gt})\big)
   + \lambda_b\big\||\mathcal{F}(\hat f)| - |\mathcal{F}(f_{gt})|\big\|_1}
   _{\text{reconstruction}}
 + \lambda_c \underbrace{\|\hat f * h - g\|_2^2}_{\text{reblur}}
 + \lambda_d \underbrace{\|\theta_E - \hat\theta_E\|_2^2}_{\text{regression}},
\label{eq:loss}
\end{equation}}

with $\lambda_a = 0.5$, $\lambda_b = 0.1$, $\lambda_c = 0.1$, and
$\lambda_d = 1$ as the optimal loss weights. Pixel $L_1 = \|\hat f - f_{gt}\|_1$ is dominated
by the low frequencies that survive blurring, the structural similarity index measure (SSIM) \citep{wang2004ssim}
rewards local structure, and the Fourier term $\big\||\mathcal{F}(\hat f)| - |\mathcal{F}(f_{gt})|\big\|$  penalizes errors at the high spatial frequencies that deconvolution must restore. The reblur residual is normalized by $\langle g^2 \rangle$ and this term re-applies the known PSF convolution and penalizes
inconsistency with the observation.

\paragraph{Metrics:}~
The fractional RMSE is used as a metric for 
$\theta_E$. For the reconstruction, because the images are sparse, 
we report $L_1$ as a percentage of the flux inside a mask ($f_{gt} > 1\%$ of peak). The deconvolution metric is the high-frequency octave error ($E_{\mathrm{top}}$):
\begin{equation}
E_{\mathrm{top}}
 = \frac{\sum_{k\in\Omega} |\hat F(k) - F(k)|^2}
        {\sum_{k\in\Omega} |F(k)|^2},
\qquad
\label{eq:hf}
\end{equation}
where $\hat F$ and $F$ are the 2D Fourier transforms of $\hat f$ and $f_{gt}$ on a $\pm4''$ box about the lens centre, and $\Omega$ is the top octave of the spatial frequencies, $\tfrac12 k_{\rm Nyq} \le |k| \le k_{\rm Nyq}$ with the Nyquist frequency of the pixel grid, $k_{\rm Nyq} = 1/(2s)$. $E_{\rm top}$ informs how far the reconstruction's top-octave coefficients are from the truth's, amplitude and phase together. This metric is normalized so that 0 is perfect, emitting nothing ($\hat f = 0$) gives a value of 1, and above 1 is worse than emitting nothing. The error bars are 16--84th percentiles from over 1000 bootstrap resamples of the 200 test lenses. We compare the model performance against two references: identity, which returns the
blurred input unchanged ($\hat f = g$), any point below it means the reconstruction is closer to the truth than the input was, and the classical Wiener deconvolution with the known PSF and regularization constant swept from $10^{1}$--$10^{-14}$.
\section{Results}
\label{sec:results}

\subsection{Deconvolution and regression across surveys}
\label{sec:invariance}

\begin{figure}[t]
\centering
\includegraphics[width=0.85\textwidth]{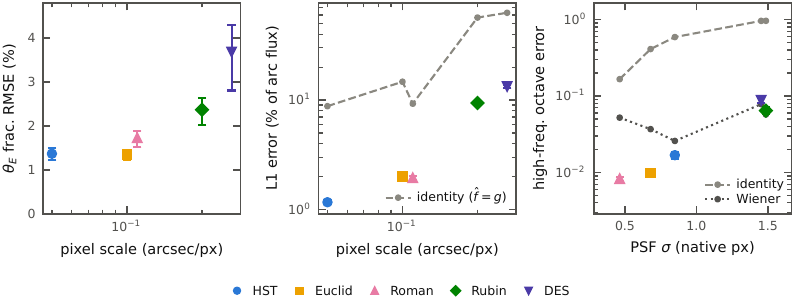}
\caption{Cross-survey performance on the test dataset. (a) Fractional
$\theta_E$ RMSE against pixel scale. (b) Masked $L_1$ as a percentage of arc
flux, with \texttt{identity} (dashed) for reference. (c) High-frequency octave error
against PSF $\sigma$ in native pixels, with \texttt{identity} (dashed) and
Wiener (dotted); $1.0$ is the emit-nothing level.}
\label{fig:invariance}
\end{figure}

Figure~\ref{fig:hero} illustrates the reconstruction capability of the model on noise-free data across the five surveys. While the pixel scale and PSF blurring is different across the surveys, the underlying signal is the same. All the images are normalized to the peak flux of the survey's true signal $f_{gt}$. The fine structure of the arc
is recovered at every pixel scale as shown in the reconstructed images $\hat f$. The residuals ($\hat f - f_{gt}$) are larger in the ground based surveys (Rubin and DES) which have coarser and more blurred images than the sky based surveys (HST, Euclid, and Roman). This is expected as there is less information available in the coarse images to being with. In Fig.~\ref{fig:invariance}, we show the model performance on the five surveys using the metrics discussed in Sec.~\ref{sec:methods}. From Fig.~\ref{fig:invariance}a, we see that the Einstein radius can be inferred within $1.3$--$3.7\%$ of the true value and the error increases with blurring. At Rubin and DES the PSF FWHM  is comparable to the width of the arcs (Table~\ref{tab:surveys}), so the arc is smeared over a few coarse pixels and are less sharply constrained than in the space-based images. In Fig.~\ref{fig:invariance}b, the masked $L_1$ error ($\|\hat f - f_{gt}\|_1$) is $4.7$--$7.5\times$ below that of the identity baseline (returning the blurred input unchanged), indicating that the model recovers the sharp flux distribution than reproducing its input. Similar to Figure~\ref{fig:hero}, the masked $L_1$ error increases from HST to DES. In Fig.~\ref{fig:invariance}c we compare the high-frequency octave error ($E_{top}$) across surveys and also with the references: identity baseline and Wiener deconvolution. The error is $10.9$--$41.7\times$ below that of the identity baseline.  
The network performs better than Wiener deconvolution in the three space-based surveys whose PSF is narrower than a native pixel ($\sigma < 1$ px, Table~\ref{tab:surveys}) and the Wiener filter is only slightly better at Rubin and DES. 

We evaluate the same noise-free-trained model on realistic noisy images for each of the five surveys. At Euclid, Roman, Rubin, and DES, the Einstein radius is nearly unaffected but the high-frequency octave error is increased by $1.3$--$4.0\times$ compared to the noise-free case indicating that the current model does not accurately recover the high spatial frequencies which are most affected by the noise. 
This motivates our future work: extending invariance from PSF and pixel scale to the noise, so that a simulation-trained model carries over to real, noisy observations. 

To test whether the learned representation of the noise-free model is invariant to the survey, we visualized the embeddings with UMAP projection. If the representations were fully invariant, embeddings from all the five surveys would overlap. Instead, the ground-based surveys group together but are clearly separated from the space-based surveys, so the latent space is not fully survey-invariant. The survey signal nonetheless decreases with network depth where the survey accounts for 61\% of the variance of the input tokens but only 4\% of the final embedding. The network therefore learns a shared, instrument-aware representation, in which the physics is encoded consistently across surveys while instrument information is retained. 

\begin{figure}[t]
\centering
\begin{minipage}[t]{0.42\textwidth}\centering
  {\small (a) Einstein radius}\\[2pt]
  \includegraphics[width=\linewidth]{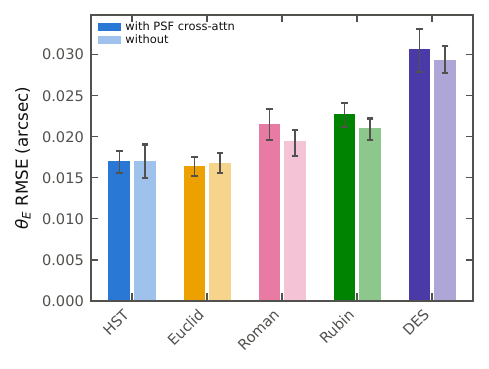}
\end{minipage}
\begin{minipage}[t]{0.42\textwidth}\centering
  {\small (b) Deconvolution}\\[2pt]
  \includegraphics[width=\linewidth]{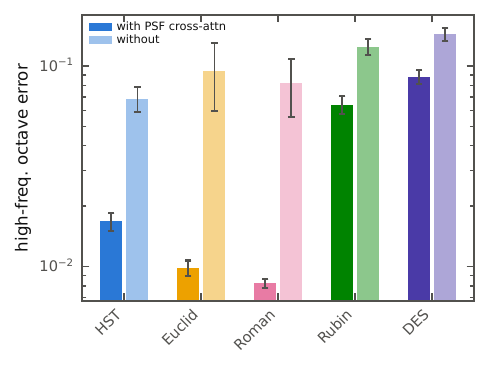}
\end{minipage}
\caption{Removing image-PSF cross-attention, with everything else held
fixed. (a) $\theta_E$ RMSE is unchanged and the ablated model is marginally better at three of the
five surveys. (b) High frequency octave error rises at every survey, from $1.6\times$ at DES to $10\times$ at Roman.}
\label{fig:crossattn}
\end{figure}

\subsection{PSF conditioning is required for deconvolution but not for $\theta_E$ estimation}
\label{sec:dissociation}

To isolate the role of PSF conditioning, we disable the image--PSF cross-attention while holding the architecture and training data fixed. Cross-attention is the only pathway by which the PSF reaches the network, so the ablated model performs both the regression and reconstruction tasks from the blurred image alone, with no knowledge of the PSF at inference. The two tasks respond very differently as shown in Fig.~\ref{fig:crossattn}.
The Einstein radius RMSE is unchanged (Fig.~\ref{fig:crossattn}a), and without cross attention, the model is in fact marginally better at three of the
five surveys. This is most likely because the structure of the arcs is not changed with PSF blurring. In contrast, deconvolution degrades at every survey (Fig.~\ref{fig:crossattn}b) without cross-attention. The high-frequency error rises by
$1.6$--$10\times$. This degradation is expected because the blur suppresses the fine structure the decoder must restore, and without the PSF information, the network can only target the PSF-marginal reconstruction, which is farther from the truth. Providing PSF explicitly improves this deblurring. 

\section{Conclusions and future work}
\label{sec:conclusions}

We present a transformer which performs strong lens deconvolution in physical coordinates and regression of Einstein radius across multiple surveys, by conditioning on the measured PSF. The Einstein radius is recovered to $1.3$--$3.7\%$ at every survey, and the high-frequency reconstruction error is lower than that of an optimally tuned Wiener filter at the three space-based surveys. The inclusion of cross-attention between the PSF and the image, the mechanism that supplies the telescope PSF to the network, improves high-frequency recovery by $1.6$--$10\times$. 

Our results demonstrate the feasibility of training a single model that generalizes across telescopes. The latent representation encodes the lens physics consistently across surveys, but is not fully survey-invariant, since the embeddings from different surveys remain distinguishable. Future work will target survey-invariance through domain adaptation \citep{ciprijanovic23,swierc2024domain,Agarwal2024}, or through contrastive learning on paired renderings of the same lens~\cite{mercaderperez26}. The model presented in this work is trained and evaluated on single-band, noise-free images with Gaussian PSFs. Future work will extend this framework to incorporate realistic observational conditions. Because our network receives the PSF as an image rather than as a parametric summary, accommodating measured, non-Gaussian PSFs requires no architectural change, and extending to multi-band data, where the PSF differs between passbands, requires widening the input embedding. Noise, by contrast, is stochastic, and robustness to noise must therefore be learned through methods such as domain adaptation or nuisance-randomized distillation \citep{puli2022ood,gandrakota2024robust}. A representation invariant to telescope properties could then serve as the front end for simulation-based inference \citep{cranmer2020frontier} and enabling the implementation of simulation-trained models to real observations.
\begin{ack}
\textit{Funding:} This work was produced by Fermi Forward Discovery Group, LLC under Contract No. 89243024CSC000002 with the U.S. Department of Energy, Office of Science, Office of High Energy Physics. Publisher acknowledges the U.S. Government license to provide public access under the DOE Public Access Plan DOE Public Access Plan
(\url{http://energy.gov/downloads/doe-public-access-plan}).

The work of S.J. was supported by the DOE Grant No. 0000258217 (FNAL 21-25).

We gratefully acknowledge the support of the NSF-Simons
AI Institute for the Sky (SkAI) via grants NSF AST-2421845
and Simons Foundation MPS-AI-00010513. 
\end{ack}

\bibliographystyle{unsrtnat}
\bibliography{refs}

\newpage
\input{checklist.tex}

\end{document}

%% file: checklist.tex
\section*{NeurIPS Paper Checklist}

\begin{enumerate}

\item {\bf Claims}
    \item[] Question: Do the main claims made in the abstract and introduction accurately reflect the paper's contributions and scope?
    \item[] Answer: \answerYes{} 
    \item[] Justification: In the abstract and introduction we clearly mention the scope of the analysis and the limitations. We present the same in the paper.

\item {\bf Limitations}
    \item[] Question: Does the paper discuss the limitations of the work performed by the authors?
    \item[] Answer: \answerYes{} 
    \item[] Justification: We present the limitations and how we plan to address them in future work in Section 4. We also present the inference results when our current model trained on noise-free data is applied on noisy data in Section 3.1

\item {\bf Theory assumptions and proofs}
    \item[] Question: For each theoretical result, does the paper provide the full set of assumptions and a complete (and correct) proof?
    \item[] Answer: \answerNA{} 
    \item[] Justification: The paper does not include any new theoretical result.

    \item {\bf Experimental result reproducibility}
    \item[] Question: Does the paper fully disclose all the information needed to reproduce the main experimental results of the paper to the extent that it affects the main claims and/or conclusions of the paper (regardless of whether the code and data are provided or not)?
    \item[] Answer: \answerYes{} 
    \item[] Justification: We describe the data and architecture in Section 2. We also provide github code.

\item {\bf Open access to data and code}
    \item[] Question: Does the paper provide open access to the data and code, with sufficient instructions to faithfully reproduce the main experimental results, as described in supplemental material?
    \item[] Answer: \answerYes{} 
    \item[] Justification: We provide the github code which includes also data generation.

\item {\bf Experimental setting/details}
    \item[] Question: Does the paper specify all the training and test details (e.g., data splits, hyperparameters, how they were chosen, type of optimizer) necessary to understand the results?
    \item[] Answer: \answerYes{} 
    \item[] Justification: We provide the training, test data details and also relevant architecture parameters in Section 2.

\item {\bf Experiment statistical significance}
    \item[] Question: Does the paper report error bars suitably and correctly defined or other appropriate information about the statistical significance of the experiments?
    \item[] Answer: \answerYes{} 
    \item[] Justification: The error bars and residuals are presented in the plots and also mentioned in the Section 2.

\item {\bf Experiments compute resources}
    \item[] Question: For each experiment, does the paper provide sufficient information on the computer resources (type of compute workers, memory, time of execution) needed to reproduce the experiments?
    \item[] Answer: \answerYes{} 
    \item[] Justification: We provide time taken for training on GPU in Section 2 along with architecture information.
    
\item {\bf Code of ethics}
    \item[] Question: Does the research conducted in the paper conform, in every respect, with the NeurIPS Code of Ethics \url{https://neurips.cc/public/EthicsGuidelines}?
    \item[] Answer: \answerYes{} 
    \item[] Justification: We provide details on the dataset and the code.

\item {\bf Broader impacts}
    \item[] Question: Does the paper discuss both potential positive societal impacts and negative societal impacts of the work performed?
    \item[] Answer: \answerNA{} 
    \item[] Justification: This analysis is specific to astronomy data and hence, does not have societal impact.
    
\item {\bf Safeguards}
    \item[] Question: Does the paper describe safeguards that have been put in place for responsible release of data or models that have a high risk for misuse (e.g., pre-trained language models, image generators, or scraped datasets)?
    \item[] Answer: \answerNA{} 
    \item[] Justification:  This work does not use pre-trained language models, image generators, or
scraped datasets, and hence poses no such risks.

\item {\bf Licenses for existing assets}
    \item[] Question: Are the creators or original owners of assets (e.g., code, data, models), used in the paper, properly credited and are the license and terms of use explicitly mentioned and properly respected?
    \item[] Answer: \answerYes{} 
    \item[] Justification: License is mentioned in the code.

\item {\bf New assets}
    \item[] Question: Are new assets introduced in the paper well documented and is the documentation provided alongside the assets?
    \item[] Answer: \answerYes{} 
    \item[] Justification: All details about the code we produced were documented and clearly presented
in the manuscript.

\item {\bf Crowdsourcing and research with human subjects}
    \item[] Question: For crowdsourcing experiments and research with human subjects, does the paper include the full text of instructions given to participants and screenshots, if applicable, as well as details about compensation (if any)? 
    \item[] Answer: \answerNA{} 
    \item[] Justification: Crowdsourcing was not used in this work.

\item {\bf Institutional review board (IRB) approvals or equivalent for research with human subjects}
    \item[] Question: Does the paper describe potential risks incurred by study participants, whether such risks were disclosed to the subjects, and whether Institutional Review Board (IRB) approvals (or an equivalent approval/review based on the requirements of your country or institution) were obtained?
    \item[] Answer: \answerNA{} 
    \item[] Justification: This work does not involve human subjects or study participants, and therefore
IRB approval or equivalent ethical review is not applicable.

\item {\bf Declaration of LLM usage}
    \item[] Question: Does the paper describe the usage of LLMs if it is an important, original, or non-standard component of the core methods in this research? Note that if the LLM is used only for writing, editing, or formatting purposes and does \emph{not} impact the core methodology, scientific rigor, or originality of the research, declaration is not required.
    \item[] Answer: \answerNA{} 
    \item[] Justification: LLMs are used for code development and editing the text of the manuscript. The research idea and conceptualization is provided by the authors.

\end{enumerate}

%% file: refs.bib
@ARTICLE{su2021roformer,
       author = {{Su}, Jianlin and {Lu}, Yu and {Pan}, Shengfeng and {Murtadha}, Ahmed and {Wen}, Bo and {Liu}, Yunfeng},
        title = "{RoFormer: Enhanced Transformer with Rotary Position Embedding}",
      journal = {arXiv e-prints},
         year = 2021,
        month = apr,
          eid = {arXiv:2104.09864},
        pages = {arXiv:2104.09864},
          doi = {10.48550/arXiv.2104.09864},
archivePrefix = {arXiv},
       eprint = {2104.09864},
 primaryClass = {cs.CL},
       adsurl = {https://ui.adsabs.harvard.edu/abs/2021arXiv210409864S}
}

@ARTICLE{puli2022ood,
       author = {{Puli}, Aahlad and {Joshi}, Nitish and {Wald}, Yoav and {He}, He and {Ranganath}, Rajesh},
        title = "{Nuisances via Negativa: Adjusting for Spurious Correlations via Data Augmentation}",
      journal = {arXiv e-prints},
         year = 2022,
        month = oct,
          eid = {arXiv:2210.01302},
        pages = {arXiv:2210.01302},
          doi = {10.48550/arXiv.2210.01302},
archivePrefix = {arXiv},
       eprint = {2210.01302},
 primaryClass = {cs.LG},
       adsurl = {https://ui.adsabs.harvard.edu/abs/2022arXiv221001302P}
}

@ARTICLE{gandrakota2024robust,
       author = {{Gandrakota}, Abhijith and {Zhang}, Lily H. and {Puli}, Aahlad and {Cranmer}, Kyle and {Ngadiuba}, Jennifer and {Ranganath}, Rajesh and {Tran}, Nhan},
        title = "{Robust anomaly detection for particle physics using multi-background representation learning}",
      journal = {Machine Learning: Science and Technology},
         year = 2024,
        month = sep,
       volume = {5},
       number = {3},
          eid = {035082},
        pages = {035082},
          doi = {10.1088/2632-2153/ad780c},
archivePrefix = {arXiv},
       eprint = {2401.08777},
 primaryClass = {hep-ex},
       adsurl = {https://ui.adsabs.harvard.edu/abs/2024MLS&T...5c5082G}
}

@ARTICLE{birrer2018lenstronomy,
       author = {{Birrer}, Simon and {Amara}, Adam},
        title = "{lenstronomy: Multi-purpose gravitational lens modelling software package}",
      journal = {Physics of the Dark Universe},
         year = 2018,
        month = dec,
       volume = {22},
        pages = {189-201},
          doi = {10.1016/j.dark.2018.11.002},
archivePrefix = {arXiv},
       eprint = {1803.09746},
 primaryClass = {astro-ph.CO},
       adsurl = {https://ui.adsabs.harvard.edu/abs/2018PDU....22..189B}
}

@book{wiener1949extrapolation,
  title     = {Extrapolation, Interpolation, and Smoothing of Stationary Time
               Series},
  author    = {Wiener, Norbert},
  publisher = {MIT Press},
  year      = {1949}
}

@ARTICLE{wang2004ssim,
       author = {{Wang}, Zhou and {Bovik}, Alan Conrad and {Sheikh}, Hamid Rahim and {Simoncelli}, Eero P.},
        title = "{Image Quality Assessment: From Error Visibility to Structural Similarity}",
      journal = {IEEE Transactions on Image Processing},
         year = 2004,
        month = jan,
       volume = {13},
       number = {4},
        pages = {600-612},
          doi = {10.1109/TIP.2003.819861},
       adsurl = {https://ui.adsabs.harvard.edu/abs/2004ITIP...13..600W}
}

@ARTICLE{cranmer2020frontier,
       author = {{Cranmer}, Kyle and {Brehmer}, Johann and {Louppe}, Gilles},
        title = "{The frontier of simulation-based inference}",
      journal = {Proceedings of the National Academy of Science},
         year = 2020,
        month = dec,
       volume = {117},
       number = {48},
        pages = {30055-30062},
          doi = {10.1073/pnas.1912789117},
archivePrefix = {arXiv},
       eprint = {1911.01429},
 primaryClass = {stat.ML},
       adsurl = {https://ui.adsabs.harvard.edu/abs/2020PNAS..11730055C}
}

@ARTICLE{swierc2024domain,
       author = {{Swierc}, Paxson and {Tamargo-Arizmendi}, Marcos and {{\'C}iprijanovi{\'c}}, Aleksandra and {Nord}, Brian D.},
        title = "{Domain-Adaptive Neural Posterior Estimation for Strong Gravitational Lens Analysis}",
      journal = {arXiv e-prints},
         year = 2024,
        month = oct,
          eid = {arXiv:2410.16347},
        pages = {arXiv:2410.16347},
          doi = {10.48550/arXiv.2410.16347},
archivePrefix = {arXiv},
       eprint = {2410.16347},
 primaryClass = {astro-ph.IM},
       adsurl = {https://ui.adsabs.harvard.edu/abs/2024arXiv241016347S}
}

@ARTICLE{Agarwal2024,
       author = {{Agarwal}, Shrihan and {{\'C}iprijanovi{\'c}}, Aleksandra and {Nord}, Brian D.},
        title = "{Neural Network Prediction of Strong Lensing Systems with Domain Adaptation and Uncertainty Quantification}",
      journal = {arXiv e-prints},
         year = 2024,
        month = oct,
          eid = {arXiv:2411.03334},
        pages = {arXiv:2411.03334},
          doi = {10.48550/arXiv.2411.03334},
archivePrefix = {arXiv},
       eprint = {2411.03334},
 primaryClass = {astro-ph.IM},
       adsurl = {https://ui.adsabs.harvard.edu/abs/2024arXiv241103334A}
}

@ARTICLE{heo2024rope,
       author = {{Heo}, Byeongho and {Park}, Song and {Han}, Dongyoon and {Yun}, Sangdoo},
        title = "{Rotary Position Embedding for Vision Transformer}",
      journal = {arXiv e-prints},
         year = 2024,
        month = mar,
          eid = {arXiv:2403.13298},
        pages = {arXiv:2403.13298},
          doi = {10.48550/arXiv.2403.13298},
archivePrefix = {arXiv},
       eprint = {2403.13298},
 primaryClass = {cs.CV},
       adsurl = {https://ui.adsabs.harvard.edu/abs/2024arXiv240313298H}
}

@ARTICLE{sureau2020spacevariant,
       author = {{Sureau}, F. and {Lechat}, A. and {Starck}, J.-L.},
        title = "{Deep learning for a space-variant deconvolution in galaxy surveys}",
      journal = {\aap},
         year = 2020,
        month = sep,
       volume = {641},
          eid = {A67},
        pages = {A67},
          doi = {10.1051/0004-6361/201937039},
archivePrefix = {arXiv},
       eprint = {1911.00443},
 primaryClass = {astro-ph.IM},
       adsurl = {https://ui.adsabs.harvard.edu/abs/2020A&A...641A..67S}
}

@ARTICLE{nammour2022shapenet,
       author = {{Nammour}, F. and {Akhaury}, U. and {Girard}, J.~N. and {Lanusse}, F. and {Sureau}, F. and {Ben Ali}, C. and {Starck}, J.-L.},
        title = "{ShapeNet: Shape constraint for galaxy image deconvolution}",
      journal = {\aap},
         year = 2022,
        month = jul,
       volume = {663},
          eid = {A69},
        pages = {A69},
          doi = {10.1051/0004-6361/202142626},
archivePrefix = {arXiv},
       eprint = {2203.07412},
 primaryClass = {astro-ph.IM},
       adsurl = {https://ui.adsabs.harvard.edu/abs/2022A&A...663A..69N}
}

@ARTICLE{akhaury2025multiband,
       author = {{Akhaury}, U. and {Jablonka}, P. and {Courbin}, F. and {Starck}, J.-L.},
        title = "{Joint multiband deconvolution for Euclid and Vera C. Rubin images}",
      journal = {\aap},
         year = 2025,
        month = may,
       volume = {697},
          eid = {A136},
        pages = {A136},
          doi = {10.1051/0004-6361/202452308},
archivePrefix = {arXiv},
       eprint = {2502.17177},
 primaryClass = {astro-ph.IM},
       adsurl = {https://ui.adsabs.harvard.edu/abs/2025A&A...697A.136A}
}

@ARTICLE{ivezic2019lsst,
       author = {{Ivezi{\'c}}, {\v{Z}}eljko and others},
        title = "{LSST: From Science Drivers to Reference Design and Anticipated Data Products}",
      journal = {\apj},
         year = 2019,
        month = mar,
       volume = {873},
       number = {2},
          eid = {111},
        pages = {111},
          doi = {10.3847/1538-4357/ab042c},
archivePrefix = {arXiv},
       eprint = {0805.2366},
 primaryClass = {astro-ph},
       adsurl = {https://ui.adsabs.harvard.edu/abs/2019ApJ...873..111I}
}

@ARTICLE{des2016overview,
       author = {{Dark Energy Survey Collaboration}},
        title = "{The Dark Energy Survey: more than dark energy - an overview}",
      journal = {\mnras},
         year = 2016,
        month = aug,
       volume = {460},
       number = {2},
        pages = {1270-1299},
          doi = {10.1093/mnras/stw641},
archivePrefix = {arXiv},
       eprint = {1601.00329},
 primaryClass = {astro-ph.CO},
       adsurl = {https://ui.adsabs.harvard.edu/abs/2016MNRAS.460.1270D}
}

@ARTICLE{euclid2025overview,
       author = {{Euclid Collaboration}},
        title = "{Euclid: I. Overview of the Euclid mission}",
      journal = {\aap},
         year = 2025,
        month = may,
       volume = {697},
          eid = {A1},
        pages = {A1},
          doi = {10.1051/0004-6361/202450810},
archivePrefix = {arXiv},
       eprint = {2405.13491},
 primaryClass = {astro-ph.CO},
       adsurl = {https://ui.adsabs.harvard.edu/abs/2025A&A...697A...1E}
}

@INPROCEEDINGS{schlieder24,
       author = {{Schlieder}, Joshua E. and others},
        title = "{Survey science with the Nancy Grace Roman Space Telescope Wide Field Instrument}",
    booktitle = {Space Telescopes and Instrumentation 2024: Optical, Infrared, and Millimeter Wave},
         year = 2024,
       editor = {{Coyle}, Laura E. and {Matsuura}, Shuji and {Perrin}, Marshall D.},
       series = {Society of Photo-Optical Instrumentation Engineers (SPIE) Conference Series},
       volume = {13092},
        month = aug,
          eid = {130920S},
        pages = {130920S},
          doi = {10.1117/12.3020622},
       adsurl = {https://ui.adsabs.harvard.edu/abs/2024SPIE13092E..0SS}
}

@ARTICLE{loshchilov2019adamw,
       author = {{Loshchilov}, Ilya and {Hutter}, Frank},
        title = "{Decoupled Weight Decay Regularization}",
      journal = {arXiv e-prints},
         year = 2017,
        month = nov,
          eid = {arXiv:1711.05101},
        pages = {arXiv:1711.05101},
          doi = {10.48550/arXiv.1711.05101},
archivePrefix = {arXiv},
       eprint = {1711.05101},
 primaryClass = {cs.LG},
       adsurl = {https://ui.adsabs.harvard.edu/abs/2017arXiv171105101L}
}

@ARTICLE{treu2010strong,
       author = {{Treu}, Tommaso},
        title = "{Strong Lensing by Galaxies}",
      journal = {\araa},
         year = 2010,
        month = sep,
       volume = {48},
        pages = {87-125},
          doi = {10.1146/annurev-astro-081309-130924},
archivePrefix = {arXiv},
       eprint = {1003.5567},
 primaryClass = {astro-ph.CO},
       adsurl = {https://ui.adsabs.harvard.edu/abs/2010ARA&A..48...87T}
}

@ARTICLE{richardson1972,
       author = {{Richardson}, William Hadley},
        title = "{Bayesian-Based Iterative Method of Image Restoration}",
      journal = {Journal of the Optical Society of America (1917-1983)},
         year = 1972,
        month = jan,
       volume = {62},
       number = {1},
        pages = {55},
          doi = {10.1364/JOSA.62.000055},
       adsurl = {https://ui.adsabs.harvard.edu/abs/1972JOSA...62...55R}
}

@ARTICLE{lucy1974,
       author = {{Lucy}, L.~B.},
        title = "{An iterative technique for the rectification of observed distributions}",
      journal = {\aj},
         year = 1974,
        month = jun,
       volume = {79},
        pages = {745},
          doi = {10.1086/111605},
       adsurl = {https://ui.adsabs.harvard.edu/abs/1974AJ.....79..745L}
}

@ARTICLE{ciprijanovic23,
       author = {{{\'C}iprijanovi{\'c}}, A. and {Lewis}, A. and {Pedro}, K. and {Madireddy}, S. and {Nord}, B. and {Perdue}, G.~N. and {Wild}, S.~M.},
        title = "{DeepAstroUDA: semi-supervised universal domain adaptation for cross-survey galaxy morphology classification and anomaly detection}",
      journal = {Machine Learning: Science and Technology},
         year = 2023,
        month = jun,
       volume = {4},
       number = {2},
          eid = {025013},
        pages = {025013},
          doi = {10.1088/2632-2153/acca5f},
archivePrefix = {arXiv},
       eprint = {2302.02005},
 primaryClass = {astro-ph.GA},
       adsurl = {https://ui.adsabs.harvard.edu/abs/2023MLS&T...4b5013C}
}

@ARTICLE{ramzi23,
       author = {{Ramzi}, Zaccharie and {Michalewicz}, Kevin and {Starck}, Jean-Luc and {Moreau}, Thomas and {Ciuciu}, Philippe},
        title = "{Wavelets in the Deep Learning Era}",
      journal = {Journal of Mathematical Imaging and Vision},
         year = 2023,
        month = jan,
       volume = {65},
       number = {1},
        pages = {240-251},
          doi = {10.1007/s10851-022-01123-w},
       adsurl = {https://ui.adsabs.harvard.edu/abs/2023JMIV...65..240R}
}

@ARTICLE{akhaury22,
       author = {{Akhaury}, Utsav and {Starck}, Jean-Luc and {Jablonka}, Pascale and {Courbin}, Fr{\'e}d{\'e}ric and {Michalewicz}, Kevin},
        title = "{Deep Learning-based galaxy image deconvolution}",
      journal = {Frontiers in Astronomy and Space Sciences},
         year = 2022,
        month = nov,
       volume = {9},
          eid = {357},
        pages = {357},
          doi = {10.3389/fspas.2022.1001043},
archivePrefix = {arXiv},
       eprint = {2211.09597},
 primaryClass = {astro-ph.IM},
       adsurl = {https://ui.adsabs.harvard.edu/abs/2022FrASS...901043A}
}

@ARTICLE{ni24,
       author = {{Ni}, Shulei and {Qiu}, Yisheng and {Chen}, Yunchun and {Song}, Zihao and {Chen}, Hao and {Jiang}, Xuejian and {Chen}, Huaxi},
        title = "{PI-AstroDeconv: A Physics-Informed Unsupervised Learning Method for Astronomical Image Deconvolution}",
      journal = {arXiv e-prints},
         year = 2024,
        month = mar,
          eid = {arXiv:2403.01692},
        pages = {arXiv:2403.01692},
          doi = {10.48550/arXiv.2403.01692},
archivePrefix = {arXiv},
       eprint = {2403.01692},
 primaryClass = {astro-ph.IM},
       adsurl = {https://ui.adsabs.harvard.edu/abs/2024arXiv240301692N}
}

@ARTICLE{patel25,
       author = {{Patel}, Aakash and {Zhang}, Tianqing and {Avestruz}, Camille and {Regier}, Jeffrey and {The LSST Dark Energy Science Collaboration}},
        title = "{Neural Posterior Estimation for Cataloging Astronomical Images with Spatially Varying Backgrounds and Point Spread Functions}",
      journal = {\aj},
         year = 2025,
        month = sep,
       volume = {170},
       number = {3},
          eid = {155},
        pages = {155},
          doi = {10.3847/1538-3881/adef32},
archivePrefix = {arXiv},
       eprint = {2503.00156},
 primaryClass = {astro-ph.IM},
       adsurl = {https://ui.adsabs.harvard.edu/abs/2025AJ....170..155P}
}

@INPROCEEDINGS{hst24,
       author = {{Wiseman}, Jennifer},
        title = "{The Hubble Space Telescope: A New Era of Discovery}",
    booktitle = {32nd General Assembly International Union (IAUGA 2024)},
         year = 2024,
        month = aug,
          eid = {3038},
        pages = {3038},
       adsurl = {https://ui.adsabs.harvard.edu/abs/2024IAUGA..32P3038W}
}

@ARTICLE{mercaderperez26,
       author = {{Mercader-Perez}, Pablo and {Cuesta-Lazaro}, Carolina and {Muthukrishna}, Daniel and {Audenaert}, Jeroen and {Villar}, V. Ashley and {Hogg}, David W. and {Huertas-Company}, Marc and {Freeman}, William T.},
        title = "{Learning What's Real: Disentangling Signal and Measurement Artifacts in Multi-Sensor Data, with Applications to Astrophysics}",
      journal = {arXiv e-prints},
         year = 2026,
        month = apr,
          eid = {arXiv:2604.09787},
        pages = {arXiv:2604.09787},
          doi = {10.48550/arXiv.2604.09787},
archivePrefix = {arXiv},
       eprint = {2604.09787},
 primaryClass = {astro-ph.IM},
       adsurl = {https://scixplorer.org/abs/2026arXiv260409787M}
}

@ARTICLE{adam25,
       author = {{Adam}, Alexandre and {Stone}, Connor and {Bottrell}, Connor and {Legin}, Ronan and {Hezaveh}, Yashar and {Perreaul-Levasseur}, Laurence},
        title = "{Echoes in the Noise: Posterior Samples of Faint Galaxy Surface Brightness Profiles with Score-based Likelihoods and Priors}",
      journal = {\aj},
         year = 2025,
        month = may,
       volume = {169},
       number = {5},
          eid = {254},
        pages = {254},
          doi = {10.3847/1538-3881/adb039},
archivePrefix = {arXiv},
       eprint = {2311.18002},
 primaryClass = {astro-ph.IM},
       adsurl = {https://scixplorer.org/abs/2025AJ....169..254A}
}

@ARTICLE{legin25,
       author = {{Legin}, Ronan and {Stone}, Connor and {Adam}, Alexandre and {Barco}, Gabriel Missael and {Coogan}, Adam and {Malkin}, Nikolay and {Perreault-Levasseur}, Laurence and {Hezaveh}, Yashar},
        title = "{Mind the Information Gap: Unveiling Detailed Morphologies of z 0.5-1.0 Galaxies with SLACS Strong Lenses and Data-Driven Analysis}",
      journal = {arXiv e-prints},
         year = 2025,
        month = nov,
          eid = {arXiv:2511.19595},
        pages = {arXiv:2511.19595},
          doi = {10.48550/arXiv.2511.19595},
archivePrefix = {arXiv},
       eprint = {2511.19595},
 primaryClass = {astro-ph.GA},
       adsurl = {https://scixplorer.org/abs/2025arXiv251119595L}
}
